\documentclass{vgtc}                          

\usepackage{enumitem}
\setlist[itemize]{leftmargin=*}
\setlist[enumerate]{leftmargin=*}

\usepackage[dvipsnames]{xcolor}
\definecolor{manniBgColor}{HTML}{f0f3f3}
\definecolor{codeColor}{HTML}{0B5A8C}
\definecolor{linkColor}{HTML}{3498db}
\definecolor{lineNumberColor}{HTML}{999999}

\usepackage[frozencache,cachedir=.]{minted}
\setminted[ts-lexer.py:TypeScriptLexer -x]{
  numbersep=5pt,
  linenos,
  fontsize=\footnotesize
}
\newminted{TypeScript}{
  linenos,
  fontsize=\footnotesize,
}
\newminted[tscodeblock]{ts-lexer.py:TypeScriptLexer -x}{
  numbersep=5pt,
  linenos,
  fontsize=\footnotesize
}
\newminted[tscodeblock2]{ts-lexer.py:TypeScriptLexer -x}{
  numbersep=5pt,
  linenos,
  fontsize=\footnotesize,
  bgcolor=manniBgColor
}
\newminted[jsxcodeblock]{jsx-lexer.py:JsxLexer -x}{
  numbersep=5pt,
  linenos,
  fontsize=\footnotesize
}
\newminted[jsxcodeblock2]{jsx-lexer.py:JsxLexer -x}{
  numbersep=5pt,
  linenos,
  fontsize=\footnotesize,
  bgcolor=manniBgColor
}

\usepackage{lmodern}

\newcommand{\link}[2]{{\textcolor{linkColor}{\urlstyle{sf}\href{#1}{#2}}}}

\newcommand{\code}[1]{\textcolor{codeColor}{\texttt{\footnotesize{#1}}}}

\ifpdf
  \pdfoutput=1\relax                   
  \usepackage{graphicx}                
  \DeclareGraphicsExtensions{.pdf,.png,.jpg,.jpeg} 
\else
  \usepackage{graphicx}                
  \DeclareGraphicsExtensions{.eps}     
\fi%

\graphicspath{{figures/}{pictures/}{images/}{./}} 

\usepackage{microtype}                 
\PassOptionsToPackage{warn}{textcomp}  
\usepackage{textcomp}                  
\usepackage{mathptmx}                  
\usepackage{times}                     
\usepackage{cite}                      
\usepackage{tabu}                      
\usepackage{booktabs}                  

\onlineid{1133}

\vgtccategory{Research}

\vgtcinsertpkg

\title{Encodable: Configurable Grammar for Visualization Components}

\author{Krist Wongsuphasawat\thanks{e-mail: krist.w@airbnb.com}}
\affiliation{\scriptsize Airbnb, Inc.}

\teaser{
 \makebox[460pt]{
 \includegraphics[width=520pt]{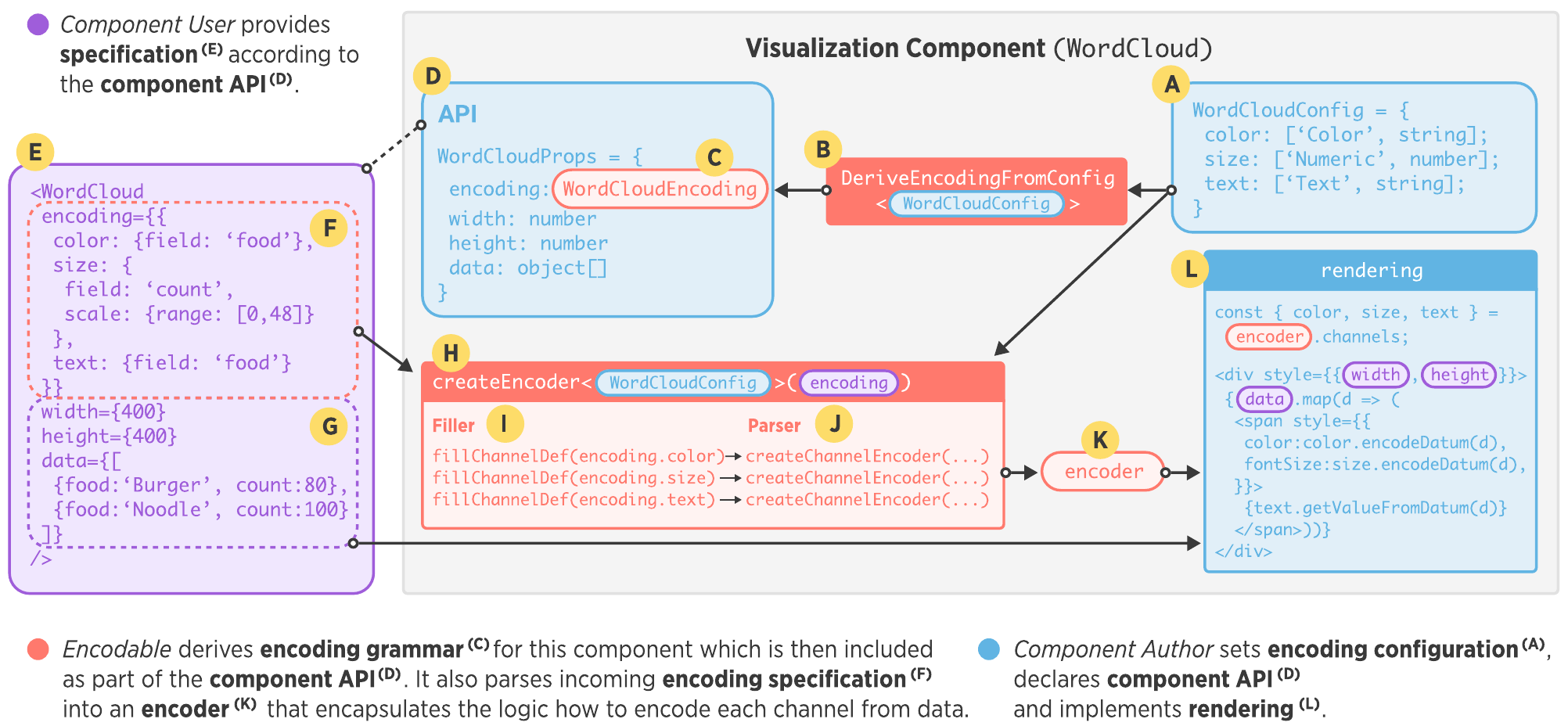}
 }
 \caption{
    The architecture of a WordCloud component built with the help of \emph{Encodable}. The colors show the roles of the component author (blue), the \emph{Encodable} library (orange) and the component user (purple). See Section~\ref{sec:solution} and~\ref{sec:demo} for more details.
 }
 \label{fig:arch}
}

\abstract{
There are so many libraries of visualization components nowadays with their APIs often different from one another.
Could these components be more similar, both in terms of the APIs and common functionalities?
For someone who is developing a new visualization component, how should the API look like?
This work drew inspiration from visualization grammar, decoupled the grammar from its rendering engine and adapted it into a configurable grammar for individual components called \emph{Encodable}. 
Encodable helps component authors define grammar for their components, 
and parse encoding specifications from users into utility functions for the implementation.
This paper explains the grammar design and demonstrates how to build components with it.

} 

\CCScatlist{
  Information visualization, systems, toolkits, API design, reusable visualization, visualization component
}

\begin{document}


\firstsection{Introduction}

\maketitle

\emph{The Grammar of Graphics} (GoG)~\cite{Wilkinson2013} introduced the idea of a single language which could express all visualizations, rather than thinking about visualizations as a catalog of charts (line, bar, pie, etc.)
Visualization libraries such as \emph{ggplot2}~\cite{Wickham2010}, and later \emph{Vega-Lite}~\cite{Satyanarayan2017}, grew from this philosophy. 
This one-grammar-to-define-them-all works really well for exploratory data analysis, which expressiveness and rapid iterations are the keys, and flourishes in the statistical and scientific communities. 
Users can fluidly transform one visualization into another by adding or modifying a few expressions. 

However, the number of libraries that follow the chart-based approach, is still growing strongly in parallel. 
This includes everything from a large library with an extensive suite of chart types~\cite{Chartjs, FusionCharts, GoogleCharts, nivo} to a tiny library with a single unique visualization. 
There are a few reasons why they keep growing:
First, when a developer already have a specific chart in mind, 
picking that chart from a catalog and setting a few options is more straightforward than learning how to express the chart via the grammar. 
Second, when someone is developing a novel visualization technique, or converting a bespoke visualization into a reusable component, he/she is likely to focus on just a single component. 
Third, performance is often an important factor for application development. 
A small library that does a few things really well can be more preferred than a large library that offers many unused functionalities.

Each of these chart-based libraries defines its own API for customizing the charts. 
Most of the time, their APIs are different from one another. 
Switching libraries means learning a new API. 
For example, to specify how to get a value for \emph{x}-position from the data: 
Some libraries take a field name string. 
Some accept a lambda function \code{xAccessor}. 
Some require each data entry to have a field named \code{x}.
They also often do not offer the same amount of common functionalities. 
For instance, some libraries support logarithmic scale, while others do not.
In some cases, even components within the same library have this discrepancy. 
For component users, are there ways these components can converge in the future to have similar APIs and common functionalities? 
For component authors, if someone wants to develop a new bespoke component or library, is there a recommended way to define its API?
From studying \emph{Vega-Lite} grammar, an idea came to mind. 
What if we build components that have APIs similar to it, but can handle the rendering ourselves?
Instead of having a grammar that can define all graphics tightly coupled with the rendering engine that transforms that grammar into actual visualizations, 
what if we decouple the grammar from the rendering engine and make it shareable among multiple components? 
Each component then can configure the shared grammar to define its scope, use that subset of grammar as its API and handle the rendering independently. 

Expanded from this idea, this work introduces
\textbf{\emph{Encodable}, a configurable grammar for encoding a component with data}. With this, component authors can: (a) \emph{Define and customize encoding grammar} for each component which conforms to the shared grammar  (b) \emph{Validate encoding specification from component users} according to the defined grammar (c) \emph{Parse encoding specification} into useful utility functions for implementing the component. 

The rest of the paper is organized as follows: 
The next section reviews relevant work, followed by an explanation of goals and requirements. 
Then the solution is described in Section~\ref{sec:solution}, accompanied by a demonstration and a brief discussion before the conclusion.




\section{Related Work}

There are many ways to create a visualization. The \textbf{programmatic approaches}, mainly on the web, can be grouped as follows:

\textbf{A) Graphics Manipulation:}
\emph{Processing}~\cite{Reas2003, Reas2005} and others~\cite{p5js, Raphael} let a developer draw or interact with visual elements directly.
They have the maximum level of expressiveness and in return require the most effort to produce the same visualizations. 

\textbf{B) Low-level Composition:}
\emph{D3}~\cite{Bostock2011} learns from the early approaches~\cite{Heer2005, Heer2010, Bostock2009} and introduces low-level building blocks, such as selection, scales, formatting, etc. 
It leverages the common standards such as SVG instead of defining all constructs by itself.
\emph{vx}~\cite{vx} bridges \emph{D3} and SVG for \emph{React}~\cite{React} framework. 
Visualizations can be created from very flexible combinations of these building blocks.

\textbf{C) Visualization Grammar:} 
Heavily inspired by the Grammar of Graphics~\cite{Wilkinson2013}, there is no concept of chart type. 
Developers learn how to express the visualizations they desire in the given grammar, i.e. a domain-specific language provided by each library that describes how to transform and encode data into visual marks and their properties~\cite{Wickham2006, Wickham2010, Satyanarayan2014, Satyanarayan2016, Satyanarayan2017, Vanderplas2018, Park2018, G2}.
The most famous one is \emph{ggplot2}~\cite{Wickham2006, Wickham2010} which dominates the R and data science communities.
\emph{Vega}~\cite{Satyanarayan2014} let users describe visualizations in JSON, and generate interactive views using either HTML5 Canvas or SVG.
\emph{Vega-Lite}~\cite{Satyanarayan2017} provides a higher-level grammar equivalent to \emph{ggplot2} level with interactions.


\textbf{D) High-level Composition:}
Similar to the convention of \emph{MS Excel}~\cite{Excel}, this group uses \emph{series} to abstract a group of graphic elements that encode data. 
For example, bars in a cartesian coordinate system form a series. 
More complex combinations such as candlestick, bullet or other chart types can also be abstracted as a series. 
The data and options are often mixed within the series definition.
\emph{ECharts}~\cite{Li2018} and others~\cite{Highcharts, Plotly} employ the all-in-one JSON option to declare a visualization.
Many libraries such as \emph{Victory}~\cite{Victory} and others \cite{React-Vis, data-ui, Semiotic}  provide similar level of abstraction in \emph{React} syntax, such as \code{<XYPlot>}, \code{<CandleStickSeries>}, or \code{<XAxis>}, that can be composed into the desired visualizations.

\textbf{E) Chart Templates:}
\emph{Google Charts}~\cite{GoogleCharts} and others~\cite{nivo, FusionCharts, JIT, Chartjs, Recharts} let developers choose a chart type from its catalog, prepare data in the specified format and plug them together. 
Some libraries provide multiple levels of abstraction. For instance, \emph{G2Plot}~\cite{G2Plot} provides chart templates on top of \emph{G2}~\cite{G2} grammar.

Encodable was designed to complement these approaches. 
It does not render the output and therefore cannot create a visualization by itself alone. 
Instead, it bridges the gap between the component authors and users. 
A component author uses \emph{Encodable} to define the component API, uses it again to parse the users' specification into an \code{Encoder}, then choose from the approaches A-D, or even E under the hood for rendering (Section~\ref{sec:solution}).
The resulting component fits into the chart templates (E) level. 

There are also some \textbf{GUI approaches} for creating visualizations with relevant concepts.
\emph{Encodable} is similar in spirit to \emph{Data-Driven Guides}~\cite{Kim2017} and the followings~\cite{Liu2018, Ren2019} which let users pick visual properties from any arbitrary shape and encode them with data.

\section{Goals \& Requirements}


This project aims to provide the following convenience:

\textbf{Component authors}, who create reusable components, should be able to create a component with \emph{encoding grammar} that conforms to this \emph{Encodable} grammar, with little effort required to make the component support the grammar. 

\textbf{Component users}, who use the reusable components, should benefit from the consistent encoding grammar across components and standardized features even though the components are from different authors.
To avoid mistakes when providing an \emph{encoding specification (spec)} for a component, users should also receive syntax verification that the spec is grammatically correct.

The goals above are broken down into the following requirements: 

\begin{itemize}
\setlength\itemsep{0.05em}
    \item \textbf{R1: Provide a configurable grammar for encoding a component with data.} The component author can customize grammar $G$ to be tailored for component $C$ as $G(C)$. $G(C)$ is still a subset of $G$ and ensures consistency across different components even though they are implemented by different component authors, e.g. $G(C_1), G(C_2), ..., G(C_n) \subset G$
    \item \textbf{R2: Handle specification parsing for the component author.} Parse the specification $\{ S \in G(C) \}$ into something that helps with the component implementation. This will also reduce the inconsistencies due to implementation of the parser.
    \item \textbf{R3: Provide mechanism to verify specification from the component users.} Learning a new grammar can take time and mistakes are inevitable. Immediate feedback when coding is very valuable to reduce mistakes from providing invalid specifications.
    \item \textbf{R4: The library should be \emph{lightweight}.} For this utility to be a dependency of any reusable component, it should not be so large that no one wants to import.
\end{itemize}

\section{Proposed Solution}
\label{sec:solution}

A grammar and parser was written in TypeScript (TS), which is a strict syntactical superset of JavaScript (JS) that adds static typing and transcompiles to JS. 
By using TS, the grammar (\textbf{R1}) can be defined as type definitions and utilize static type checking to compare incoming specifications against the type definitions. 
This will ensure that the component users have specified the specifications that are grammatically correct (\textbf{R3}).
The overall architecture of \emph{Encodable} can be seen in Fig.~\ref{fig:arch}.
Encodable components assume the datasets are in tabular format such as:

\begin{minted}[fontsize=\footnotesize]{json}
[ {"kind":"Cat", "count":9}, {"kind":"Dog", "count":11} ]
\end{minted}

The code snippets in this paper are simplified for explanation purposes and may omit some details for brevity. Please see the supplementary materials for more details or repository (\link{https://github.com/kristw/encodable}{github.com/kristw/encodable}) for the full and latest implementation.

\subsection{The Grammar}

The first principle of \emph{Encodable} is each visualization has one or more channels to encode data, such as \emph{color}, \emph{x}, \emph{y}, etc. 
For example, a simple word cloud component has \emph{size} and \emph{text} channels.
If there is a grammar to describe what \emph{size} and \emph{text} can be,
one can describe how to encode this word cloud component with the given data based on these two channels.
Hence, in its simplest form, \emph{Encodable} grammar is defined as key-value pairs of channel names and their definitions. 

\subsubsection{Channel Definition}
~\label{sec:channel_def}

This work was heavily inspired by \emph{Vega-Lite}, which includes channel definitions as part of its grammar. 
Its grammar is also pure JSON and can be serialized into a simple text file.
In \emph{Vega-Lite}, this is how to encode a bar chart that shows number of each animal:

\vspace*{-0.25em}
\begin{tscodeblock}
const vegaLiteBarSpec = { "mark": "bar",
  "encoding": {
    "x": {"field": "kind", "type": "ordinal"},
    "y": {"field": "count", "type": "quantitative"} }};
\end{tscodeblock}

In the example above, the first channel name is \emph{x} and its channel definition is \code{\{"field": "kind", "type": "ordinal"\}}, telling the rendering engine to encode the \code{kind} field for $x$-position and \code{count} field for $y$-position, or bar height.
\emph{Encodable} adopts a subset of grammar from \emph{Vega-Lite} for channel definition (\code{ChannelDef}). 

\vspace*{-0.25em}
\begin{tscodeblock}
interface ValueDef {
  value: number | string | boolean | Date | null; }
interface FieldDef {
  field: string;  format?: string;  title?: string; }
interface ScaleFieldDef extends FieldDef { 
  type: 'quantitative'|'ordinal'|'temporal'|'nominal' 
  scale?: ScaleDef;  /* See Supp. Materials */ }
interface PositionFieldDef extends ScaleFieldDef { 
  axis?: AxisDef;    /* See Supp. Materials */ }
type ChannelDef = 
  ValueDef | FieldDef | ScaleFieldDef | PositionFieldDef;
\end{tscodeblock}

According to the grammar defined above, a channel definition can be one of the followings: 

(a) \textbf{Fixed value} (\code{ValueDef}) 
-- such as making \emph{color} of text in a word cloud always red.

(b) \textbf{Dynamic value based on a field in the data} (\code{FieldDef}) 
-- such as using the field \code{kind} for each word in word cloud.

(c) \textbf{Dynamic value with scale} (\code{ScaleFieldDef}) 
-- Many channels use scale to map input value into output such as mapping \code{kind} into \emph{color}, \code{count} into \emph{fontSize}. 
Inside the \code{scale} field, the component users can define how they want to customize the scale.
The \code{type} field in channel definition will help the \emph{filler} choose the appropriate \code{scale} or \code{format} when not specified. E.g., a \code{quantitative} field uses a linear scale with number formatter by default while a \code{temporal} field uses a time scale with time formatter by default. The two scale types handle ticks and domain rounding differently.

(d) \textbf{Dynamic value with scale and axis} (\code{PositionFieldDef}) 
-- Channels such as \emph{x} or \emph{y} can optionally include definition for axes.

\vspace*{-0.25em}
\begin{tscodeblock}
const color:ValueDef = { value: 'red' };
const text:FieldDef = { field: 'kind' };
const color:ScaleFieldDef = {
  type: 'nominal', field: 'kind',
  scale: { type: 'ordinal', range: ['pink', 'blue']} };
const fontSize:ScaleFieldDef = {
  type: 'quantitative', field: 'count', 
  scale: { range: [0, 36]} };
const y:PositionFieldDef = {
  type: 'quantitative', field: 'count',
  scale: { nice: true }, axis: { orient: 'left' } };
\end{tscodeblock}

\subsubsection{Define Component-specific Channels}
\label{sec:encoding_config}

At the time of this writing, \emph{Vega-Lite} has 35 channels (\emph{x}, \emph{y}, \emph{color}, etc.) 
Even so, there are still edge cases that are beyond these fixed set of channels. 
For example, if the developer is trying to encode data into font-family, there is no such channel in \emph{Vega-Lite} and therefore you cannot use it.
So a fixed number of channels does not sound like a good idea.
Earlier in Section~\ref{sec:channel_def}, \emph{Encodable} grammar is defined broadly as a key-value object (\code{Encoding}) with key being channel name and value being channel definition. 
This basically allows unlimited number of channels. 

\vspace*{-0.5em}
\mint[fontsize=\footnotesize]{TypeScript}|interface Encoding { [channelName: string]: ChannelDef }|

However, this is too ambiguous and problematic. \code{channelName} can be any \code{string}. 
There is nothing to enforce component users to specify the correct channel names, which basically violates \textbf{R3}. 
Users may specify channel \code{color} when there is no such channel in the component.
Also each channel may support only a subset of the \code{ChannelDef} type. 
E.g., a text channel does not care about axis or scale and should only be \code{ValueDef} or \code{FieldDef}. 

Therefore, the second principle of \emph{Encodable} is the component authors can define channel names and definitions specific to their components via a configuration below. 

\vspace*{-0.25em}
\begin{tscodeblock}
type ChannelType = 'X'|'Y'|'Numeric'|'Category'|'Color'|'Text';
type Output = number | string | boolean | null;
interface EncodingConfig {
  [name: string]: [ChannelType, Output, 'multiple'?];  
}
\end{tscodeblock}

Component authors must list their channel names with their types, expected output type, and whether it can take multiple (array of) definitions (such as a tooltip channel can accept multiple fields to be displayed). 
For example, to create a word cloud component that can be encoded by color and font size and accept multiple fields for tooltip, the component author will write this configuration (Fig.~\ref{fig:arch}-A) and derive the encoding grammar from the config (Fig.~\ref{fig:arch}-B).

\vspace*{-0.25em}
\begin{tscodeblock}
import { DeriveEncoding } from 'encodable';
interface WordCloudConfig {
  color: ['Color', string];
  fontSize: ['Numeric', number];
  text: ['Text', string];
  tooltip: ['Text', string, 'multiple'] }
type WordCloudEncoding = DeriveEncoding<WordCloudConfig>;
\end{tscodeblock}

In \code{DeriveEncoding} (Fig.~\ref{fig:arch}-B), each \code{ChannelType} in the config is mapped to an appropriate subset of channel definition as follows:

\begin{center}
\vspace*{-0.5em}
\footnotesize
 \begin{tabular}{c c} 
 Channel Type & Channel Definition \\ [0.5ex] 
 \hline
 \code{X, Y} & \code{PositionFieldDef|ValueDef}  \\ 
 \hline
 \code{Numeric, Category, Color} & \code{ScaleFieldDef|ValueDef}  \\ 
 \hline
 \code{Text} & \code{FieldDef|ValueDef}  \\ 
 \hline
\end{tabular}
\end{center}

\code{X} and \code{Y} channel types represent x- and y- positions. 
\code{Numeric} channel type means a numeric attribute, e.g., size, opacity.
\code{Category} channel type defines a categorical attribute, e.g., visibility, shape.
\code{Color} channel type defines a color attribute, e.g., fill, stroke.
\code{Text} channel type defines a plain text attribute, e.g. tooltip, label.
The grammar \code{WordCloudEncoding} derived from the \code{WordCloudConfig} is equivalent to the manually-defined \code{WordCloudEncoding} below. 
However, the extra information in config that a channel is a \code{Color} type, not an ordinary \code{Category} will be useful during parsing, which the manual one cannot capture.

\vspace*{-0.25em}
\begin{tscodeblock}
type WordCloudEncoding = {
  color: ValueDef | ScaleFieldDef<string>;
  fontSize: ValueDef | ScaleFieldDef<number>;
  text: ValueDef | FieldDef<string>;
  tooltip: (ValueDef | FieldDef<string>)[]; /* array */}
\end{tscodeblock}

\subsection{The Encoder}

\emph{Encodable} takes encoding config (Fig.~\ref{fig:arch}-A) from the author and encoding specification from the user (Fig.~\ref{fig:arch}-F), and parses it into an \code{Encoder} (Fig.~\ref{fig:arch}-K) that encapsulates the logic how to encode each channel from data (\textbf{R2}).
During parsing, each channel definition is parsed separately. 
Since many fields are optional, the \emph{filler} (Fig.~\ref{fig:arch}-I) will expand the incoming definition into a completed definition via smart defaults and inference.
After that, each channel definition is parsed into a \code{ChannelEncoder} (Fig.~\ref{fig:arch}-J), which is a utility class that provides several useful functions, such as:
\code{encodeDatum(datum)} which converts input datum into output value for that channel and \code{getValueFromDatum(datum)} which returns the raw field value from input datum, or fixed value, for that channel. 
All \code{ChannelEncoder} instances are nested under an \code{Encoder} instance and referred to by \code{encoder.channels[channelName]}. 
The author then can use the \code{Encoder} and these \code{ChannelEncoder} to help with the rendering (Fig.~\ref{fig:arch}-L) of the visualization. 




The \emph{Encodable} library, at the time of this writing, is 25.2kB (minified), which is relatively small (\textbf{R4}). 
In comparison, \emph{Vega-Lite} is 237.1kB, \emph{G2} is 414.9kB and \emph{Echarts} is 817kB.

\section{Demonstration}
\label{sec:demo}

The code below demonstrates how to implement the rendering logic of the word cloud component. 
It is the completed version of Fig.~\ref{fig:arch}-L.
Line 5 defines encoding grammar of this component as the \code{WordCloudEncoding} defined earlier in Section~\ref{sec:encoding_config}. 
Line 8 parse incoming encoding specification into an \code{Encoder}.
Line 9 sets the domain from data. 
For example, if \emph{color} is based on the field \code{count}, this call will set the domain of the color channel to \code{[min(count), max(count)]}.
This single call applies the same operation to all channels.
When rendering the HTML \code{<span>} (line 11-16), the three \code{ChannelEncoder}: \code{size}, \code{color} and \code{text} are used to computed the output for each channel from each datum. 
Whether the color is a fixed value, comes from what field, uses a quantized scale or other scales, the component author does not need to know because these logic are encapsulated within the \code{ChannelEncoder}. 
Although the example is based on \emph{React}, \emph{Encodable} is independent from \emph{React} and can work with other frameworks, or even plain JS.

\makeatletter
\expandafter\def\csname PYGdefault@tok@err\endcsname{\def\PYGdefault@bc##1{##1}}
\makeatother
\begin{jsxcodeblock}
import { createEncoder } from 'encodable';
export function WordCloud({
  encoding, width, height, data
}: {
  encoding: WordCloudEncoding; 
  width: number; height: number; data: object[];
) {
  const encoder = createEncoder<WordCloudConfig>(encoding);
  encoder.setDomainFromDataset(data);
  return (<div style={{ width, height }}>
    {data.map(d => (<span style={{
      color: encoder.channels.color.encodeDatum(d),
      fontSize: encoder.channels.size.encodeDatum(d),
    }}>
      {text.getValueFromDatum(d)}
    </span>))}
  </div>);
}
\end{jsxcodeblock}


\emph{Encodable} greatly reduces the overhead in adding or removing encoding channels. 
Adding a \emph{fontWeight} channel to the word cloud above requires only two small changes: (1) Add the \code{fontWeight} channel to \code{WordCloudConfig}. (2) Add \code{fontWeight} property to the \code{<span>} on line 14 above.
With this lighter overhead, the author can develop a prototype with the core visual elements first and decide on adding, changing or removing encoding channels later. 
This also helps with standardizing and converting an existing component or custom graphics into a reusable component by substituting hard-coded value or old code with an encoding channel.

\begin{figure}
\vspace*{-1.5em}
\centering
\includegraphics[width=200pt]{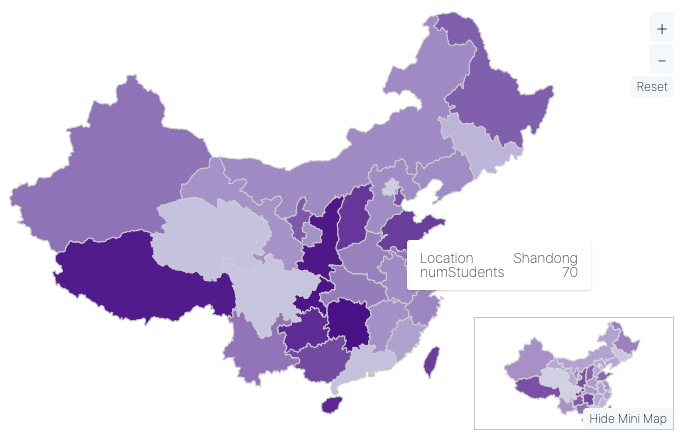}
\caption{
A China map component with a mini-map. The SVG rendering and zoom/pan interactions are implemented with \emph{React}. The visual encoding is handled by \emph{Encodable}. 
The component author includes the following channels: \emph{location}, \emph{fill}, \emph{stroke}, \emph{texture} and \emph{tooltip}. 
}
\label{fig:map}
\vspace*{-1em}
\end{figure}

\begin{figure}
\centering
\includegraphics[width=230pt]{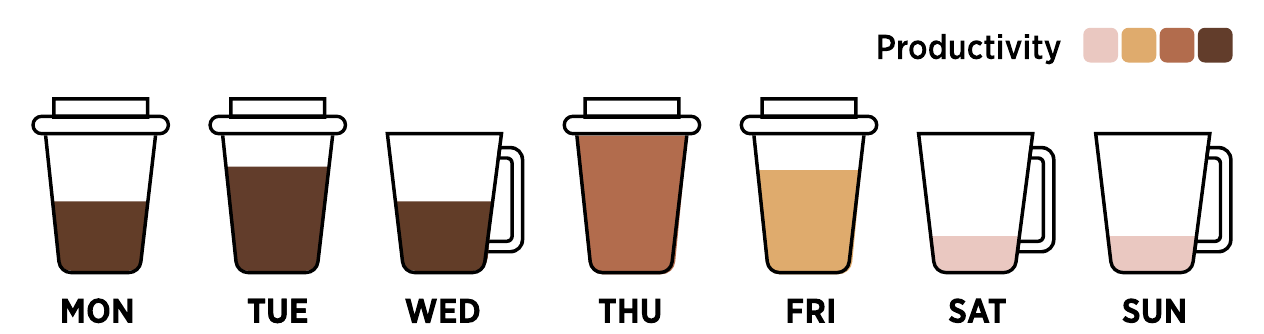}
\caption{
A bespoke visualization component of coffee cups. The visual encoding is handled by \emph{Encodable} 
with the following channels: \emph{drinkLevel}, \emph{label}, \emph{drinkColor} and \emph{useToGoCup}. 
}
\label{fig:coffee}
\end{figure}

While the word cloud example is easier to explain, it does not represent the full potential of this work. 
\emph{Encodable} is capable of enabling more complex components. 
In the China map example (Fig.~\ref{fig:map}), the author builds a traditional map component that has several encoding channels, then a user encodes \emph{fill} channel with \code{numStudents}, resulting in a choropleth map with a sequential color scale. 
The coffee chart (Fig.~\ref{fig:coffee}) allows different parts of a coffee cup to be encoded by data. 
A user then encodes the cups to represent his productivity and coffee consumption over the week.

\begin{jsxcodeblock}
<ChinaMap data={data} encoding={{ 
  location: { field: 'province' },
  fill: { field: 'numStudents', type: 'quantitative' } }} />
<CoffeeChart data={productivityData} encoding={{ 
  label: { field: 'day' },
  drinkLevel: { field: 'numCoffee', type: 'quantitative' }, 
  drinkColor: { field: 'productivity', type: 'quantitative' },
  useToGoCup: { field: 'goToOffice', type: 'ordinal' } }} />
\end{jsxcodeblock}

For real applications, \emph{Encodable} was used to build several components (scatter plot, box plot, line chart, map, etc.) for the open-source project \emph{Apache Superset}. The components are now part of the official application release. The package \code{encodable} is also available on npm registry with 3,000 weekly downloads.

\section{Conclusion and Future Work}
\label{sec:conclusion}

This work envisions a world where visualization components from different authors can have consistent APIs and behavior. 
This does not limit to traditional charts, but also applies to bespoke visualizations.
Inspired by \emph{Vega-Lite}, a new configurable grammar independent from rendering called \emph{Encodable} is introduced.
It lets a component author declare a grammar for encoding channels of his/her component, which looks like a subset of \emph{Vega-Lite} grammar. 
To ease the implementation burden, the grammar is accompanied with a parser that parses specifications from component users into utility functions to help with the rendering.
To provide feedback for the component users, the specifications can also be verified against the grammar.
The demonstration shows that it is easy to configure encoding channels and can support a broad range of components, making component authoring convenient and flexible. 

Looking ahead, there are still many things that could be added to this configurable grammar, such as legend and axis support, more features in the channel definition, etc. 
Expanding these new ideas while keeping the library lightweight will be an interesting challenge.

\acknowledgments{
Thanks to Kanit Wongsuphasawat, Dominik Moritz,
Chris Williams \& Airbnb Data Experience team for their feedback and support.
}

\bibliographystyle{abbrv-doi}

\bibliography{ms}
\end{document}